\documentclass[prx,%
 reprint,
 amsmath,amssymb,
 aps,
]{revtex4-2}

\usepackage{graphicx}
\usepackage{dcolumn}
\usepackage{bm}
\usepackage{hyperref}

\usepackage{graphicx}
\usepackage{dcolumn}
\usepackage{bm}
\usepackage{bbm}
\usepackage{soul}
\usepackage{enumitem}
\usepackage{makecell}

\usepackage{braket}
\usepackage{graphicx}
\usepackage{amsmath,verbatim,latexsym,amssymb,indentfirst,mathrsfs,mathtools,amsthm,bbm,bm,hyperref,url}

\usepackage[title,titletoc]{appendix}

\usepackage{cancel}
\usepackage{bbold}
\usepackage[dvipsnames]{xcolor}
\usepackage[normalem]{ulem}
\usepackage{physics}

\begin{document}

\preprint{APS/123-QED}

\title{Surface Morphology Assisted Trapping of Strongly Coupled Electron-on-Neon Charge States}

\author{Kaiwen Zheng}
 \affiliation{Physics Department, Washington University, Saint Louis, MO, USA, 63130.}
\author{Xingrui Song}
 \affiliation{Physics Department, Washington University, Saint Louis, MO, USA, 63130.}
\author{Kater W. Murch}%
 \email{murch@physics.wustl.edu}
\affiliation{Physics Department, Washington University, Saint Louis, MO, USA, 63130.}

\date{\today}

\begin{abstract}
Single electrons confined to a free neon surface and manipulated through the circuit quantum electrodynamics (circuit QED) architecture is a promising novel quantum computing platform. Understanding the exact physical nature of the electron-on-neon (eNe) charge states is important for realizing this platform's potential for quantum technologies. We investigate how resonator trench depth and substrate surface properties influence the formation of eNe charge states and their coupling to microwave resonators. Through experimental observation supported by modeling, we find that shallow-depth etching of the resonator features maximizes coupling strength. By comparing the trapping statistics and  surface morphology of devices with altered trench roughness, our work reveals the role of fabrication-induced surface features in the formation of strongly coupled eNe states. 

\end{abstract}

\maketitle

\begin{figure}
    \includegraphics{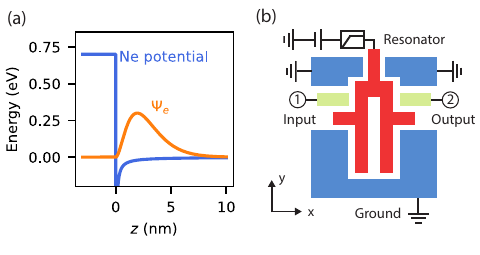}
    \caption{\textbf{Sketch of the eNe system.} \textbf{(a)} The Ne potential strongly confines the electron in the $z$ direction. Motional states in the $x$--$y$ plane can therefore couple to planar microwave circuitry. \textbf{(b)} Simplified sketch of the planar microwave device which features a voltage-biased coplanar stripline resonator.}
    \label{fig1}
\end{figure}

When electrons are brought close to the free surface of a solid Ne film  they can be spontaneously confined to a plane that is parallel to and \rm{$\sim$}2 nm above the surface. This occurs due to a combination of the attractive dielectric polarization induced on the neon surface and repulsion from its full valence shell as shown in Fig.~\ref{fig1}(a)~\cite{Cole_electron_trapping, eH2_and_eNe, Kajita_SN, Kajita_SN2}. Recent work~\cite{Dafei_Ne, Dafei_NatPhys, xinhao} has shown that a Ne film deposited on a circuit QED device can lead to confined eNe charge states that are strongly coupled to a coplanar stripline resonator. The demonstrated eNe charge states, with transition energies in the GHz range, have coherence times greatly exceeding traditional semiconductor charge qubits~\cite{semiQubit} and other similar floating electron systems~\cite{WeiGuo_review, Ash_review, GeYang, Koolstra, Mikolas, Erika}. Strong coupling to a microwave resonator allows state readout and single-qubit operations through microwave drives.  These features make eNe charge states a promising novel quantum computing platform.

The eNe charge states were first believed to be single electrons confined laterally in a trap region by the electrostatic potential generated from electrodes in its vicinity, similar to that of single electrons on He~\cite{Koolstra}. 
However, the deviations of the eNe transition energies and stability from what would be predicted based on the electrode-derived Coulomb potential have motivated theoretical investigation into the role of surface morphology as a trapping mechanism ~\cite{toshi}. 
That theory suggests that trapped electron states in the GHz frequency range occur due to bumps and valleys on the neon surface. Despite qualitative alignment with experimental observations, there has not yet been experimental verification of the hypothesis. 

In this work, we compare electron trapping experiments conducted on Ne films deposited onto three types of devices. The devices share a similar planar design but different vertical geometries and surface morphologies enabled by slightly altered fabrication procedures.   When probing in the GHz frequency range, we find that devices with a rough (etched Si)  surface can easily trap strongly coupled eNe states. In contrast, a device with a smooth surface (sapphire substrate) cannot trap strongly coupled eNe states. 
These observations motivate us to propose a model for trapped eNe charge states, where relatively smooth polycrystalline Ne film conforms to much rougher irregularities on the exposed substrate surface. The combined features are of the size and shape that would result in localized eNe states in the GHz frequency range based on recent theoretical work~\cite{toshi}. Furthermore, we experimentally show and verify through modeling that a shallow etching into the microchip substrate results in eNe charge states with greater coupling strength to the resonator.

\begin{table*}
\begin{ruledtabular}
\begin{tabular}{lccccc}
\textrm{Device}&
\textrm{$\omega_\mathrm{CM}/2\pi$ (GHz)}&{\textrm{$\kappa_\mathrm{CM}/2\pi$ (kHz)}}&{\textrm{\textrm{$\omega_\mathrm{DM}/2\pi$ (GHz)}}}&{\textrm{$\kappa_\mathrm{DM}/2\pi$ (kHz)}}&{$t$ \textrm{(nm)}}\\
\colrule
Shallow Si & 4.5361 & 209.0 & 4.9251 & 154.2 & 75\\
Deep Si & 5.2202 & 705.0 & 6.1839 & 994.0 & 1100\\
Sapphire 1 & 4.5564 & 2905.9 & 5.6406 & 886.6 & 0\\
Sapphire 2 & 4.9072 & 367.8 & 5.6788 & 265.4 & 0\\
Sapphire 3 & 4.5361 & 167.2 & 4.9251 & 234.3 & 0\\
\end{tabular}
\end{ruledtabular}
\caption{\label{table1}%
Parameters of all devices used in this study, including the resonance frequencies of the common (differential) mode resonances, $\omega_\mathrm{CM}$ ($\omega_\mathrm{DM}$), their respective linewidths, $\kappa_\mathrm{CM}$ ($\kappa_\mathrm{DM}$), and trench depths, $t$.
}
\end{table*}

Previously demonstrated eNe devices utilized coplanar stripline resonators to strongly couple to eNe charge states.
These devices featured a defined trap region where the two traces of the resonator bend to provide electric field along both $x$ and $y$ axes, confining an electron~\cite{Koolstra, Dafei_Ne, Dafei_NatPhys, xinhao}. Several gate electrodes were also placed close to the trap region to provide additional fields to define the charge qubit. In this study, we utilize a simplified device design consisting of straight resonator traces at the voltage anti-node. A DC bias voltage is only applied to the resonator electrodes. This design eliminates a defined trap region. A simplified drawing of the device architecture used in this study is shown in Fig.~\ref{fig1}(b). We fabricate these devices on either intrinsic (100) Si or c-plane sapphire wafers. For the Si devices, a 140 nm Nb film is deposited onto the Si substrate using electron beam evaporation immediately after piranha and buffered oxide etch treatments. For the sapphire devices, a 200 nm Nb film is sputtered onto the substrate immediately after piranha treatment \footnote{The Nb films on sapphire wafers were purchased from STAR Cryoelectronics.}. Both types of film have similar elongated grain structures and are therefore considered to have similar surface properties~\cite{DariaNb, passivation, andrewHouckNb}. The films are patterned using electron beam lithography with ZEP 520A resist, and the structures are transferred using reactive ion etching (RIE) with a mixture of $\rm{SF}_6$ and Ar gas. The etching process removes Nb that is not protected by the resist, and for the Si device can be used to further etch into the substrate forming a trench region between metallic electrodes. By varying the etching time, we produce trenches into the Si with different depths. In one Si ``shallow'' device we etch $t\sim 75$ nm into the substrate. This is close to the minimal value required to prevent accidental shorts between metallic structures. In the ``deep" trench device, we  over-etch the substrate by \rm{$t\sim$}1100 nm. This is close to the maximum depth required to prevent delamination of the Nb film. The etching recipe does not attack the sapphire substrate and therefore results in a trench depth $t\sim 0$ nm. The relevant parameters of the shallow and deep Si devices, as well as three sapphire devices are listed in Table~\ref{table1}.

\begin{figure*}
\centering
    \includegraphics{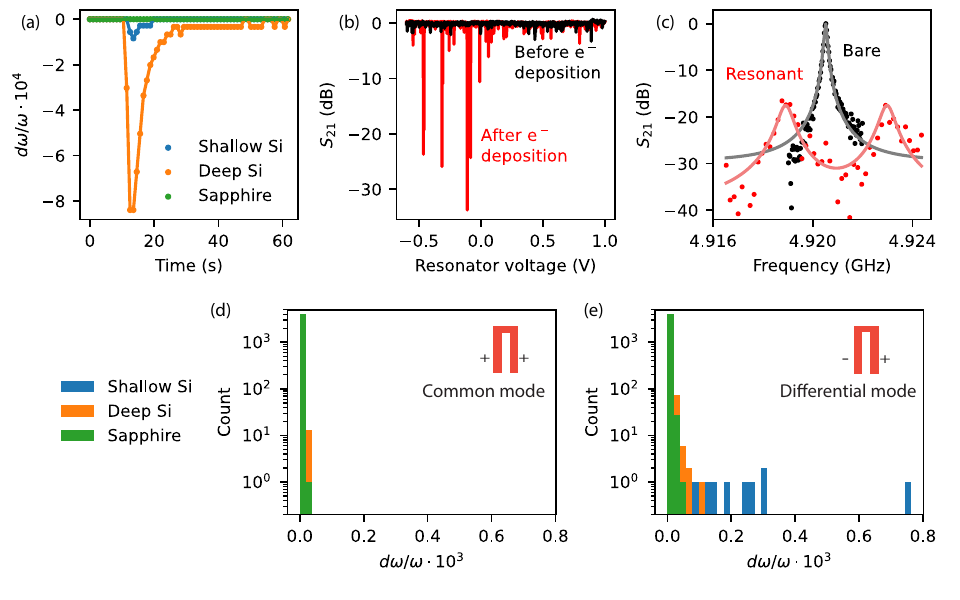}
    \caption{\textbf{Electron-on-Ne experiments across different devices.} \textbf{(a)} Real-time resonance frequency shift of the DM of the shallow Si, deep Si, and a sapphire device during electron deposition. \textbf{(b)} $S_{21}$ transmission of the DM versus resonator voltage of the Ne-covered shallow Si device before and after electron deposition. \textbf{(c)} $S_{21}$ transmission of the DM versus frequency at (red) and away (black) from an eNe avoided crossing. \textbf{(d, e)} Histograms of the fractional frequency shifts measured as the resonator voltage is stepped between $-0.6$ and $1$~V in $200\ \mu$V steps.  }
    \label{fig2}
\end{figure*}

The fabricated devices are mounted onto a printed circuit board housed in a hermetic sample cell. We deposit the Ne films by injecting controlled amounts of Ne gas into the sample cell held at a temperature $T\sim 27$~K, slightly above the triple point of Ne. We then slowly cool the sample cell to $T\sim 3$~K over a period of 10 hours. Then, the sample is further cooled to a temperature of  $T\sim 8$~mK. Similar to previous work, the Ne films are kept at a thickness of roughly 10~nm, estimated by the frequency shift of the microwave resonator due to the changes in its characteristic capacitance from the Ne dielectric. Electrons are deposited by sending pulsed voltages of various lengths and duty cycles through a tungsten filament mounted inside the sample cell close to the microchip. Typical electron emission events cause the sample package to heat up temporarily by 1--400~mK depending on the pulse area. As shown in Fig.~\ref{fig2}(a), the resonator mode frequencies decrease during the electron deposition and relax back to their original frequency. This frequency shift is likely due to the generation of thermal quasi-particles in the superconducting structures  \cite{McRae_review}. We observe that for the same energy deposition to the sample package, the deep Si device tends to have the greatest frequency response, and the sapphire devices have the smallest response. This is consistent with the higher thermal conductivity of sapphire at cryogenic temperatures \cite{sapphireAttenuator}. 

After the sample package returns to its base temperature, we tune the DC voltage applied to the resonator to search for strongly coupled eNe charge states.  The resonator structure features common mode (CM) and differential mode (DM) resonances, which can be probed by a common microwave feedline. We monitor the transmission scattering matrix element \rm{$S_{21}$} at these resonance frequencies while stepping the resonator voltage. Figure~\ref{fig2}(b) shows $S_{21}$ of the DM for two representative scans of the Ne-covered shallow Si device before and after electron deposition. Before electron deposition, we observe small transmission dips of $<5$~dB corresponding to two-level-system (TLS) defects. The applied electric field can tune the transition energies of these TLS defects into resonance with the DM, reducing the transmission~\cite{muller_TLS, TLS_dipole, Lisenfeld_TLS}. 

In contrast, after electron deposition, we observe a large number of transmission dips as the resonator voltage is tuned. We associate these with eNe charge states that are strongly coupled to the DM resonance. The resulting avoided crossing causes a significant drop in \rm{$S_{21}$} at the probe frequency. An example of avoided crossing is shown in Fig.~\ref{fig2}(c), where the DM resonance peak splits due to resonant strong coupling with an eNe state. This splitting of \rm{$2g=2\pi\times(4.04~\mathrm{MHz})$ results in a $\sim 30$~dB drop in \rm{$S_{21}$}}. This coupling rate corresponds to an estimated electric dipole moment of \rm{$\sim 30$} e\r{A} \footnote{As the electrons on the Ne surface are strongly confined in the $z$ direction with the first excitation energy of a few meV, we assume that the eNe states interact with the resonator through the electric field in the $x$ axis. Assuming that the eNe state is at the voltage antinode of the DM, we estimate the electric dipole moment using Eq.~\ref{eq1}.}, which is about an order of magnitude larger than the typical value for TLS defects observed in superconducting circuit experiments~\cite{TLS_dipole, Lisenfeld_TLS, muller_TLS}. Although the transmission dips appear only in negative applied voltages in the scan shown in Fig.~\ref{fig2}(b), we also observe dips at positive applied voltages in other scans. To quantify the coupling strength of eNe states across different devices, we use the resonator linewidth to convert the transmission dips into estimated frequency shifts. The extracted fractional resonator frequency shifts are displayed for both CM and DM in Figs.~\ref{fig2}(d, e). The statistics of the fractional frequency shifts are used to evaluate each device's ability to trap eNe states with strong coupling to the microwave resonator. From these measurements, we highlight three features that we will discuss in detail in the following. First, across all three devices we observe no eNe states with large \rm{$d\omega/\omega$} on the CM resonance. Second, we observe eNe charge states with significantly greater \rm{$d\omega/\omega$} on the DM resonance of shallow Si device than the deep Si device. Finally, we observe no states with observable \rm{$d\omega/\omega$} on the sapphire devices.  

\begin{figure}
\begin{center}
    \includegraphics{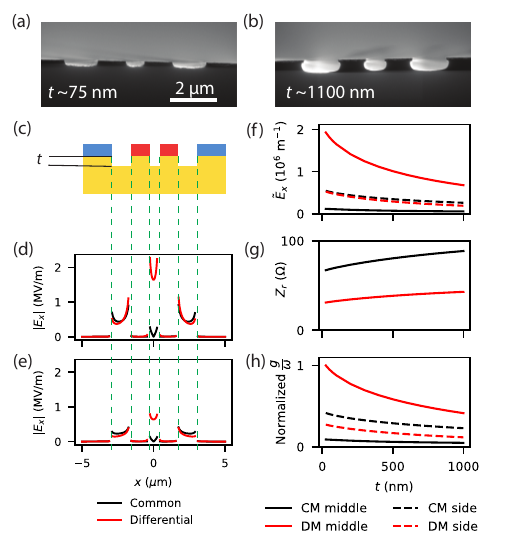}
\end{center}
    \caption{\textbf{Cross section analysis of a coplanar stripline resonator.} \textbf{(a, b)} Secondary electron scanning electron microscopy image of the cleaved cross section of \textbf{(a)} the shallow trench Si device, and \textbf{(b)} the deep Si trench device. \textbf{(c)} Sketch of the resonator cross section with an over-etched trench depth $t$. \textbf{(d, e)} Simulated lateral electric field strength \rm{$|E_x|$} \textbf{(d)} $t$ = 50 nm, and \textbf{(e)} $t$ = 1000 nm. \textbf{(f)} Simulated \rm{$\tilde{E_x}$} inside the middle and side trenches for the CM and DM resonances.  \textbf{(g)} Characteristic impedance \rm{$Z_r$} for the two resonance modes. \textbf{(h)} Normalized coupling strength  for eNe states trapped in the trench regions. }
    \label{fig3}
\end{figure}

To interpret our findings for the two Si devices, we model the cross section of the coplanar stripline resonator shown in Figs.~\ref{fig3}(a, b) with idealistic anisotropic over-etched trenches of various depth using finite element simulation [Fig.~\ref{fig3}(c)]. We set both traces of the resonator at a voltage of 1~V in magnitude and with the same or opposite polarity to mimic the CM and DM, respectively. Figs.~\ref{fig3}(d, e) show the magnitude of the lateral component of the simulated electric field \rm{$|E_x|$}, 5~nm above the exposed metal or trench surface. The two modes have vanishing \rm{$|E_{x}|$} on top of the superconducting electrodes and comparable \rm{$|E_{x}|$} in the side trench regions (between the traces and the ground plane). We note two features that will impact the coupling strength of the eNe charge states to the resonator modes. First, in the middle trench (region between the two traces), the DM has a \rm{$|E_x|$} that is significantly larger than for the CM. Second, \rm{$|E_x|$} in the trenches decreases with increasing trench depth. 

For an eNe state interacting with a resonator mode, the maximal coupling strength is \rm{$g=\mu\cdot E_{ZPF}$}, where \rm{$\mu$} is the electric dipole moment of the trapped eNe state transition and \rm{$E_{ZPF}$} is the electrical field at the voltage anti-node caused by the zero point fluctuation of the mode~\cite{SchusterThesis, cQED_review}. For the \rm{$\lambda$}/4 resonators used in this study, \rm{$E_{ZPF}=\tilde{E_x}\cdot V_{ZPF}$}, where \rm{$\tilde{E_x}=(\frac{1}{w}\cdot\int_0^w|E_x|dx)/(1 \mathrm{V})$}, is the average \rm{$|E_{x}|$} within a trench of width $w$ per volt applied to the trace electrodes, and \rm{$V_{ZPF}$} is the anti-node voltage due to zero point fluctuations. \rm{$V_{ZPF}$} can be calculated via the characteristic impedance of the resonance mode, \rm{$Z_\mathrm{r}$}, and the resonance frequency; \rm{$V_{ZPF}=\omega\sqrt{\frac{2Z_\mathrm{r}}{\pi\hbar}}$}~\cite{cQED_review, SchusterThesis}. This leads to the ratio of the coupling strength to the resonance frequency,
\begin{equation}
    \frac{g}{\omega}=\mu \tilde{E_x}\sqrt{\frac{2Z_\mathrm{r}}{\pi\hbar}}.
    \label{eq1}
\end{equation}
From this equation, we can estimate \rm{$g/\omega$} of the resonator modes based on \rm{$\tilde{E_x}$} and \rm{$Z_\mathrm{r}$} values extracted through simulations. As shown in Figs.~\ref{fig3}(f, g, h), as $t$ increases, \rm{$\tilde{E_x}$} decreases while \rm{$Z_\mathrm{r}$} increases with a weaker dependence, resulting in an overall decreasing \rm{$g/\omega$} which we have normalized to its maximum value. Figure~\ref{fig3}(h) reveals that an eNe charge state inside the middle trench that is coupled to the DM has the largest \rm{$g/\omega$}. 
Figure~\ref{fig3}(h) reveals that the largest \rm{$g/\omega$} can be achieved for an eNe charge state located inside the middle trench region and coupled to the DM. Furthermore, $g/\omega$ is maximized for small trench depths. 

\begin{figure*}
\centering
    \includegraphics{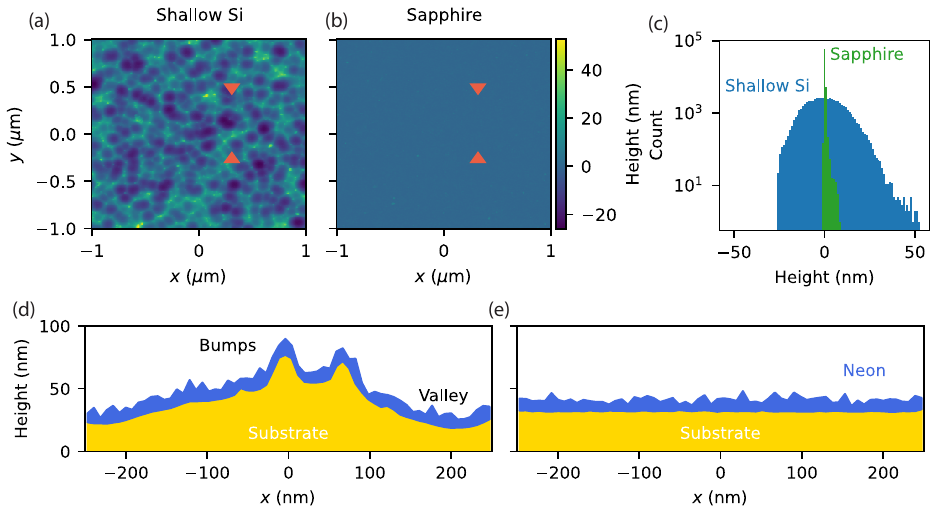}
    \caption{\textbf{Trench roughness study of eNe devices.} \textbf{(a, b)} AFM image of the exposed \textbf{(a)} Si and \textbf{(b)} sapphire surface after Nb etching. \textbf{(c)} Histogram of the AFM height data of both devices. \textbf{(d, e)} Proposed surface morphology for the Ne-coated Si and sapphire devices. The substrate profiles are taken from the AFM images at the regions indicated by arrows. The sapphire device produces a relatively smoother surface without features capable of trapping eNe states in the GHz frequency range.}
    \label{fig4}
\end{figure*}

Across three different devices fabricated on sapphire substrates, we consistently do not observe significant $d\omega/\omega$ shifts of the DM resonance after depositing electrons. 
We posit that this is due to the dramatically different surface morphology between the shallow Si and the sapphire devices revealed with atomic force microscopy (AFM) in Figs.~\ref{fig4}(a, b). We observe numerous valleys with width of \rm{$\sim$}200~nm and depth \rm{$\sim$}25~nm separated by ridges, as well as a few spikes with width \rm{$\sim$}10~nm on top of the ridges. The valleys and the spike features are also visible in the histogram plot of the AFM data shown in Fig.~\ref{fig4}(c). The different surface morphologies are a result of the fabrication process; SF\rm{$_6$}-based RIE chemistry usually results in a rough Si surface after etching due to the relatively low selectivity against Si~\cite{Pappas_SF6, mechanical_noise}. As SF\rm{$_6$} does not etch sapphire, the sapphire surface is much smoother with a root-mean-square roughness \rm{$R_q$}=0.4 nm compared to an \rm{$R_q$}=9.5 nm for the Si device.

We propose a model of the Ne surface morphology for the Si and sapphire devices in Figs.~\ref{fig4}(d, e). The substrate profile is taken from a representative line scan of the AFM data.  A thin film of Ne with mean thickness of 10~nm, thickness standard deviation of \rm{$\sqrt{10}$}~nm, and log-normal distribution of height, typical of polycrystalline thin films, is coated onto the substrates~\cite{nucleationGrowth, grainSizeThinFilm, lognormal, lognormal2}. For the Si device, despite the thickness variation in the Ne film caused by its polycrystalline \cite{rareGasSolidsV2} nature, the morphology of the surface is still dominated by the features on the substrate created during device fabrication. These bumps and valleys can have feature sizes of tens of nm. Prior theoretical predictions~\cite{toshi} have shown that features of this size can lead to eNe states with GHz transition energies. In comparison, a Ne film with the same thickness variation observed in the sapphire device produces a much smoother surface that may lack the features needed for confined eNe states with GHz transition frequencies and large dipole moments. 

Our work suggests that the trench surface morphology plays a significant role in trapping electrons through surface valleys and bumps. Although the trench roughness does not affect microwave loss and is therefore not an important parameter for fabricating superconducting qubits, devices used for controlling eNe qubits may require special attention to surface morphology. Moving forward, devices with engineered bumps and valleys with desired geometry may provide a pathway for producing eNe charge states with predictable frequencies, anharmonicities, and dipole moments---essential for scaling this novel qubit architecture into a viable quantum computing platform.

\begin{acknowledgments}
The authors thank Dafei Jin, Xinhao Li,  Toshiaki Kanai, and Johannes Pollanen for fruitful discussions. The authors acknowledge Xinyi Zhao, Zachary T. Bernard, Yinyao Shi, Sidharth Duthaluru, and Matthew K. Fowlerfinn for their assistance with the experimental setup. We thank Arpit Ranadive and Nicolas Roch for providing the traveling wave parametric amplifier (TWPA) used in this study. This work is supported by the Gordon
and Betty Moore Foundation, DOI 10.37807/
gbmf11557. The
authors acknowledge the use of facilities at the Institute of Materials
Science and Engineering in Washington University.
\end{acknowledgments}

\appendix
\section{Device Layout}
\begin{figure}
\centering
    \includegraphics{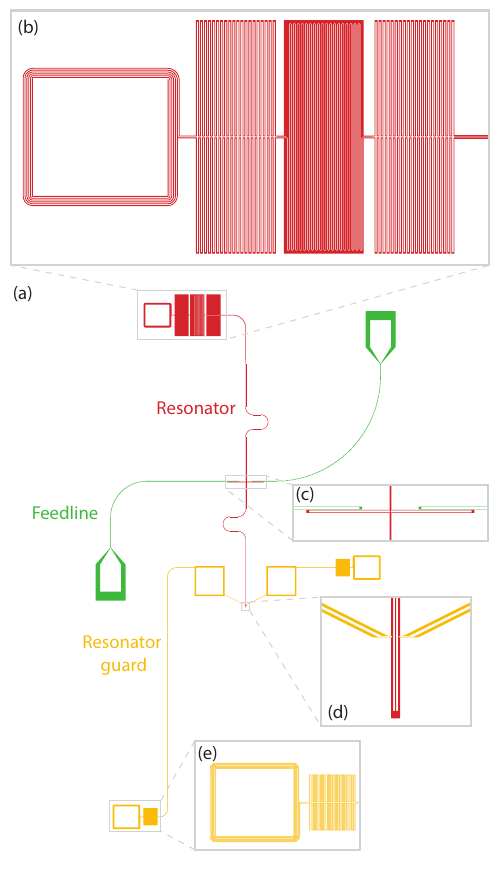}
    \caption{\textbf{Device layout used for this study.} \textbf{(a)} The chip design. \textbf{(b)} On-chip low pass filter for applying a DC bias voltage to the resonator. \textbf{(c)} The resonator is capacitively coupled to two feedlines for microwave transmission measurements. \textbf{(d)} The voltage anti-node region of the resonator. \textbf{(e)} On-chip low pass filter for the resonator guard electrodes, which remain grounded for this study.}
    \label{fig5}
\end{figure}

Figure~\ref{fig5}(a) shows the design layout of the devices used in this study. The device comprises a coplanar stripline resonator connected to a low pass filter (shown in Fig.~\ref{fig5}(b). The low pass filter allows us to add a DC bias to the resonator while maintaining a high quality factor~\cite{xiaoMi_filter, Koolstra, GeYang}. As shown in Fig.~\ref{fig5}(c), two coplanar waveguide feedlines are capacitively coupled to the resonator to probe the transmission signal. Two ``resonator guard" electrodes are placed close to the voltage anti-node of the resonator as shown in Fig.~\ref{fig5}(d). These electrodes each have an on-chip low pass filter as shown in Fig.~\ref{fig5}(e). We ground both of the resonator guard electrodes in this study. 

\section{Experimental Setup}
\begin{figure*}
\centering
    \includegraphics{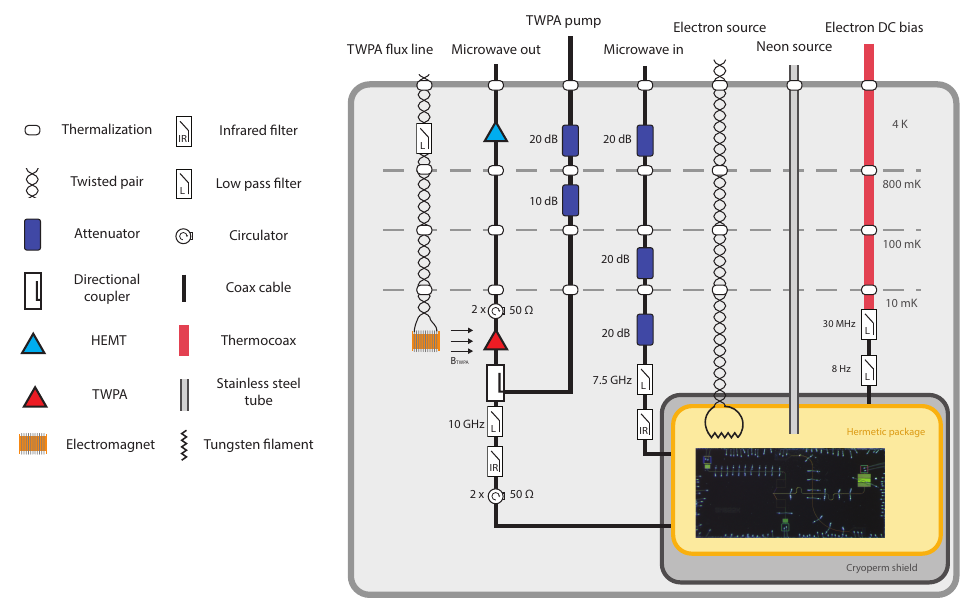}
    \caption{\textbf{Electrical and gas handling setup of the cryostat.} }
    \label{fig6}
\end{figure*}

Figure~\ref{fig6} shows the setup for this study. The chip is placed in a hermetic sample cell thermalized to the mixing chamber stage of a dilution refrigerator. A stainless steel tube for Ne deposition is only thermalized to the sample package and the 4 K stage.  A tungsten filament is fixed inside the sample package to deposit electrons. The DC bias on the resonator is fed in through a thermocoax cable that is filtered by two low pass filter stages at the mixing chamber. The respective cut off frequencies are 30 MHz and 8 Hz. The microwave probe signal is attenuated at various stages of the dilution refrigerator (total of 60 dB of attenuation, a 7.5 GHz low pass filter, and an infrared filter). The output signal is amplified by a traveling wave parametric amplifier (TWPA)~\cite{Roch} at the mixing chamber stage and a high mobility electron transistor (HEMT) amplifier at the 4 K stage.  

%

\setcounter{equation}{0}
\setcounter{figure}{0}
\setcounter{table}{0}
\setcounter{page}{1}
\setcounter{section}{0}
\makeatletter
\renewcommand{\theequation}{S\arabic{equation}}
\renewcommand{\thefigure}{S\arabic{figure}}
\renewcommand{\thetable}{S\arabic{table}}
\renewcommand{\thesection}{S\arabic{section}}
\makeatother

\end{document}